\documentstyle[preprint,aps,epsfig]{revtex}

\newcommand{\be}{\begin{eqnarray}}
\newcommand{\ee}{\end{eqnarray}}

\begin{document}

%\draft command makes pacs numbers print

\preprint{
SNUTP 98--093
}

\title{ 
Evolution equation for the heavy quark distribution function\\
in the parton model approach
}
\vspace{0.65in}

\author{
Kang Young ~Lee\thanks{kylee@ctp.snu.ac.kr }
and 
Seungwon ~Baek \thanks{swbaek@phya.snu.ac.kr} 
}
\vspace{.5in}

\address{
Center for Theoretical Physics, Seoul National University,
Seoul 151 -- 742, Korea}

\date{\today}
\maketitle
\begin{abstract}

We explore the perturbative QCD corrections to 
the inclusive semileptonic decays of $B$ mesons.
In inclusive $B$ decay processes 
the parton picture works well and the scaling feature manifests 
because the mass of $b$ quark is larger than the QCD scale.
Due to this property, the decay rate may be expressed by a single
structure function describing the light--cone distribution
of $b$ quark apart from the kinematic factor.
We derive a $q^2$ evolution equation for the distribution function,
which violates the scaling property
through the Altarelli--Parisi type approach.
The evolution equation is numerically solved 
and its phenomenological implications are also discussed.

%{\bf This is the final draft.}

\end{abstract}

\pacs{ }

% body of paper here
\narrowtext
%\tighten

\section{Introduction}

Our understanding of inclusive decays of $B$ mesons
has made great progress over the last few years.
Since the $B$ meson is heavy, the momentum transfer of the decay
is larger than the typical energy scale of hadronic bindings
$\sim \Lambda_{QCD}$ in most region of phase space
and so the light--cone dynamics dominates.
This enables us to formulate the inclusive decay precesses
in an analogous manner to that of the deep inelastic
scattering (DIS).
The light--cone dominance allows us to express the commutator
of two currents as bilocal operator of quark fields.
The Fourier transform of the bilocal operator is related
to the structure function of the decay which describe the momentum
distribution of a $b$ quark inside $B$ mesons.
Nonperturbative power corrections are encoded in the matrix elements
of the bilocal operator and also in the distribution function.
Because of asymptotic freedom of QCD,
we expect that the $B$ meson could be treated as a group of
partons neglecting interactions between quarks
following Feynman's parton picture \cite{feynman},
which leads to the simple modellings of the inclusive decay
processes \cite{paschos1,kylee1,kylee2,pas2,accmm}.

With applying the operator product expansion (OPE) to the hadronic
tensor, the leading twist contributions to the power corrections 
for the $B$ decay processes have been extensively studied 
in the framework of the heavy quark effective theory (HQET)
and the hadronic matrix elements are systematically parametrized
in terms of $1/m_b$ expansions
\cite{chay,hqet,falk}.
However the complete descriptions of the decay processes are not 
attained yet when the final state quark mass is small
because of its singular behavior at the endpoint region
where the OPE fails.
Although the scaling property due to the heaviness of the $b$ quark
makes it possible to express the decay process
by a single universal function,
this function cannot be calculated from QCD at present.
We still have unknown part of nonperturbative power corrections
as well as we need to resum the leading twist contributions
to the distribution function which becomes important at this region.
Several authors have proposed the resummation of 
the leading twist contributions and smearing to obtain a well-defined
shape function \cite{bigi1,bigi2,neubert}.
As a result we still lean on modelling the distribution function
to take into account the decay spectra at the endpoint region
\cite{aleph}.
Therefore it is interesting to analyze phenomenological models
in the framework of the HQET.
The $1/m_b$ structure of modelling the inclusive decays 
based on defining the mass of $b$ quark as
\be
m_b = \langle x \rangle m_B
\ee
are studied in the Refs. \cite{kylee1,kylee2} 
which leads to the parametrization
of the nonperturbative contributions.
Besides we can also find many attempts to the analysis
of phenomenological models based on the HQET \cite{bigi1,accmmh}.

On the other hand there are perturbative corrections which have
logarithmic singularities close to the endpoint at the quark level, 
which are regularized when integrating over the phase space 
and summing over ensembles of states.
In phenomenological and theoretical ground,
it is important to incorporate both perturbative and nonperturbative
contributions at the endpoint region.
In order to accomplish the task 
an evolution equation for the heavy quark distribution function
has been studied 
in Refs. \cite{grozin,balzereit}.

In the present paper, we  attempt to derive an evolution equation
for the distribution function in the intuitive way
introduced by Altarelli and Parisi \cite{ap} to combine 
the perturbative corrections into the parton model approach
and solve the equation.
This type of approaches avoids the formal tools of QCD but
is completely equivalent.
The phenomenological implications of the evolution are
investigated through the $q^2$-- and charged lepton energy spectra.

This paper is organized as follows: In Sec. II, we briefly review
the formalism for inclusive semileptonic decays of $B$ meson.
We show that the formulation by Jin and Paschos \cite{jp,jin1}
is equivalent to the simple parton picture.
The derivation of the evolution equation for the distribution
function of the heavy quark is given by the method of
Altarelli--Parisi in Sec. III.
The equation is solved for the moments of the distribution function
in Sec. IV and also solved numerically in Sec. V.
We obtain the evolution of the distribution function and 
investigate the phenomenologies of the decay spectra.
Finally we conclude our work in Sec. VI.

\section{Inclusive semileptonic $B$ decays}

We have the differential decay width for the process
$B \to X_q l \nu$
\be
d\Gamma = \frac{G_F^2 |V_{qb}|^2}{2 E_B} 
                       L_{\mu \nu} W^{\mu \nu} 
                               \frac{d^3 p_l}{(2\pi)^3 2 E_l}
                               \frac{d^3 p_\nu}{(2\pi)^3 2 E_\nu},
\ee
where $L_{\mu \nu}$ is the leptonic tensor and
the hadronic tensor $W^{\mu \nu}$ is defined as
\be
W^{\mu \nu} = \frac{1}{2} \sum_{X}
     \int \frac{d^3 p_{_X}}{(2\pi)^3 2E_X} 
                (2\pi)^4 \delta^4 (p_{_B} - q -p_{_X})
     \langle B | {J^\nu}^{\dagger} | X \rangle
     \langle X | J^\mu | B \rangle~ ,
\ee
which can be written in a more compact form
\be
W^{\mu \nu} = -\frac{1}{2} \int d^4x~ e^{i q \cdot x}
     \langle B | [ J^\mu (x),{J^\nu}^{\dagger}(0)] | B \rangle
\ee
by using a completeness relation $\sum | X \rangle \langle X | = 1$ 
and physical condition $q_0 > 0$.
The $B$ meson state is covariantly  normalized by
\be
     \langle B(p') | B(p) \rangle = 2 E_B (2 \pi)^3 
                        \delta^3 ({\bf p}' - {\bf p}).
\nonumber 
\ee
It would be helpful for the OPE that we relate $W^{\mu \nu}$
to the absorptive part of the forward scattering amplitude
\be
T^{\mu \nu} = -\frac{i}{2} \int d^4x~ e^{i q \cdot x}
     \langle B | T J^\mu (x){J^\nu}^{\dagger}(0) | B \rangle~,
\ee
through the optical theorem $W^{\mu \nu} = 2$ Im$T^{\mu \nu}$.

In most region of the phase space, we expect that
the dominant contribution of the integration in Eq. (4)
to $W^{\mu \nu}$ comes from 
the region close to the light--cone, $0 \le x^2 \le 1/q^2$ 
We write the commutator on the light--cone as \cite{jp,jin1}
\be
\langle B | [ J_\mu (x),{J_\nu}^{\dagger}(0)] | B \rangle
   = 2 (S_{\mu \alpha \nu \beta} - i \epsilon_{\mu \alpha \nu \beta})
     (\partial^\alpha \Delta_q(x))
     \langle B |\bar{b}(0) \gamma^\beta (1-\gamma_5) b(x)| B \rangle~,
\ee
where $S_{\mu \alpha \nu \beta}=g_{\mu \alpha} g_{\nu \beta}
             +g_{\mu \beta} g_{\nu \alpha}-g_{\mu \nu} g_{\alpha \beta}$
and $ \Delta_q(x)$ is the Pauli--Jordan function
for the produced quark $q$ defined as
\be
i\Delta_q(x) \equiv \int \frac{d^4p}{(2\pi)^3}
e^{-ip \cdot x} \epsilon (p_0) \delta (p^2 - m_q^2)~.
\nonumber
\ee
Note that Eq. (6) is Lorentz covariant and 
can be calculated in any reference frame.

Light--cone dominance enables us to keep only the leading term
in the light--cone expansion of the above reduced matrix element
and we define the quark distribution function by the Fourier
transformation of it:
\be
f(z) = \frac{1}{4\pi m_{_B}^2}
       \int d(x \cdot p_{_B})~ e^{i z x \cdot p_{_B}}
  \langle B |\bar{b}(0) p_\beta 
             \gamma^\beta (1-\gamma_5) b(x)| B \rangle |_{x^2=0}~.
\ee
After integration over $x$, we obtain the hadronic tensor
\be
W_{\mu \nu} = 
 4 \pi (S_{\mu \alpha \nu \beta} - i \epsilon_{\mu \alpha \nu \beta})
 \int dz~ f(z)~ \epsilon(z p_{_B 0} - q_0)~ \delta((z p_{_B}-q)^2 - m_q^2)
                  ~ (z p_{_B}-q)^\alpha p_{_B}^\beta~.
\ee

We consider the hadronic tensor in the light of the parton model.
Using the relation of decaying quark momentum $p_b = z p_{_B}$ 
and produced quark momentum $p_q = z p_{_B}-q$, we rewrite
the hadronic tensor in the form:
\be
W^{\mu \nu} 
%&=& \int \frac{dz}{z} f(z) \int \frac{d^3p_q}{(2\pi)^3 2E_q}
%              (2\pi)^4 \delta^4 (p_b - q -p_q)~~~~~~~~
%&&~~~~~~~~ \times~ 2~ (p_b^\mu p_q^\nu + p_b^\nu p_q^\mu 
%                                 - g^{\mu \nu} p_b \cdot p_q
%-i \epsilon^{\mu \nu \alpha \beta} p_{b \alpha} p_{q \beta})
%\nonumber \\
= \int \frac{dz}{z}~ f(z)~ \widetilde{ W}^{\mu \nu}~,
\ee
where 
\be
\widetilde{W}^{\mu \nu} =
\int \frac{d^3p_q}{(2\pi)^3 2E_q}
              (2\pi)^4 \delta^4 (p_b - q -p_q)
\cdot 2~ (p_b^\mu p_q^\nu + p_b^\nu p_q^\mu 
                                 - g^{\mu \nu} p_b \cdot p_q
-i \epsilon^{\mu \nu \alpha \beta} p_{b \alpha} p_{q \beta})
\nonumber
\ee
is the spin sum of two quark currents like the leptonic tensor. 
We find that actually the hadronic tensor at the hadronic level  
is the convolution of the hadronic tensor at the quark level and 
it leads to the convolution equation of the differential decay rate
\be
d \Gamma (B \to X_q l \nu) = \int dz~ f(z)~ d \Gamma(b \to q l \nu)~.
\ee
We find that Feynman's parton picture manifests in this approach.

From Eq. (2), we obtain the triple differential decay rate
\be
\frac{d\Gamma}{dE_l dq^2 dq_0}
= \frac{G_F^2 |V_{ub}|^2}{4\pi^3 m_{_B}}
  \frac{q_0-E_l}{\sqrt{{\bf q}^2+m_q^2}}
  \left[ f(x_+)(2 x_+ E_l m_{_B} - q^2) - (x_+ \to x_-) \right]~,
\ee
where $x_{\pm} = [(q \cdot p_{_B}) \pm
\sqrt{(q \cdot p_{_B})^2 - m_B^2 (q^2-m_q^2)}]/m_B^2$.
The contribution of the second term dependent on $x_-$
is negligible because the function $f(x)$ should be sharply peaked.
This is true after the evolution, which was checked numerically later.
If we let $m_q$ be zero and take the $B$ rest frame,
finally we obtain the triple differential decay rate
revealing the scaling behavior
\be
\frac{d\Gamma}{dE_l dq^2 dq_0}
\approx \frac{G_F^2 |V_{ub}|^2}{192\pi^3}
            \frac{2}{m_{_B}} f(x_+)
            \frac{24(q_0-E_l)(2m_bE_l-q^2)}{m_b-q_0}~,
\ee
which effectively depends upon one scaling variable $x_+$ alone 
apart from kinematic factors.
Here $m_{b}$ is as defined in Eq. (1). 
%the final state quark is a $u$-quark and so $m_q=0$.
If we consider the perturbative QCD corrections,
additional $q^2$ dependence comes in and 
the scaling behavior should be violated.
We study the evolution of the distribution function of $b$ quark
with respect to $q^2$ in the next section.

\section{Evolution equation for the distribution function}
%\label{sec:intro}

To incorporate the perturbative effects of QCD 
with our viewpoint of the parton model,
it is very suggestive to apply the method of the Altarelli--Parisi 
equation \cite{ap,roberts}
since the inclusive semileptonic decays of heavy flavours
are intimately related to DIS via channel crossing.
This method will show close contact with the parton model approach 
to inclusive $B$ decays at each step.
We derive the evolution equation for the light--cone 
distribution function of $b$ quark in this section.

We decompose the $B$ meson decay process into the subprocess as 
shown in Fig. 1 to write the amplitude
\footnote{Note that we distinguish the produced quark $q$ and
other final states $X$ in this section. Please see Fig.1 (a).
Namely we call the final states (momenta) the produced quark $q$ $(p_q)$,
the rest hadronic degrees of freedom $X$ $(p_{_X})$
and lepton pair $W$ $(q)$.}:
\be
{\cal M}(B \to q X W) = {\cal M}(B \to X b) {\cal M}(b \to q W) 
                \frac{1}{E_X+E_b-E_B} \frac{1}{2 E_b}~.
\ee
We substitute the decomposed amplitude in Eq. (2), we have
\be
d \Gamma(B \to q X W) &=& 
\frac{1}{2 E_B} |{\cal M}(B \to X b)|^2 |{\cal M}(b \to q W)|^2 
                \frac{1}{(E_X+E_b-E_B)^2} \frac{1}{(2 E_b)^2}
\nonumber \\
&& \times (2\pi)^4 \delta^4 (p_{_B} - q -p_{_X}-p_q)
                   \frac{d^3 p_{_X}}{(2\pi)^3 2E_X} 
                   \frac{d^3 p_q}{(2\pi)^3 2 E_q}
                   \frac{d^3 q}{(2\pi)^3 2 q_0}~,
\ee
and comparing this equation with Eq. (10), we obtain
\be
f(z) dz = \frac{E_b}{E_B} |{\cal M}(B \to X b)|^2 
                \frac{1}{(E_X+E_b-E_B)^2} \frac{1}{(2 E_b)^2}
                   \frac{d^3 p_{_X}}{(2\pi)^3 2E_X} ~,
\ee
where $z = E_b/E_B$.

We consider the one-gluon emission from the $b$ quark
which brings out the leading correction to the decay process
of order of $\alpha_s$.
Our strategy is to describe the effects of the one-gluon emission 
by the change of the distribution function.
Let us consider $b \to q l \nu$ process as a two-body decay
with the final state quark $q$ and the lepton pair $(l\nu) \equiv W$.
At the quark level, the differential decay rate is given by
\be
\frac{d \Gamma (b \to q W)}{d q_0} 
= \Gamma_0~ \delta(q_0-\frac{1}{2}m_b)
\equiv \frac{2 \Gamma_0}{m_B}~ \delta(\xi-\xi_+)~,
\ee
where $\xi_+ = 2 q_0/m_B$ and $q$ is taken to be a momentum of
$W$ boson (lepton pair).
At the hadronic level, we have
\be
\frac{d \Gamma (B \to  q X W)}{d q_0} 
= \int dz~ f(z)~ \frac{d \Gamma (b \to q W)}{d q_0} 
\equiv \frac{2 \Gamma_0}{m_B} f(\xi_+)~.
\ee
Thus we express the distribution function with an argument $y_+$
as the differential decay rate.
The gluon emission yields the change of the distribution function
\be
\Delta f (\xi) = \frac{m_B}{2 \Gamma_0} 
\frac{d \Gamma (B \to  q X W + g)}{d q_0}~. 
\ee
As in the Eq. (13), we decompose the gluon emission process
into three subprocesses
\be
{\cal M}(B \to  q X W + g) 
      &=& {\cal M}(B \to X b) {\cal M}(b \to q W + g) 
                \frac{1}{E_X+E_b-E_B} \frac{1}{2 E_b}~,
\\
{\cal M}(b \to q W + g) 
      &=& {\cal M}(b \to b' g) {\cal M}(b' \to q W ) 
                \frac{1}{E_g+E_{b'}-E_b} \frac{1}{2 E_{b'}}~,
\ee
as shown in Fig. 1 (b).
The differential decay rates are given by
\be
d \Gamma(B \to  q X W + g) &=& \frac{1}{2 E_B}
 |{\cal M}(B \to X b)|^2 |{\cal M}(b \to b' g)|^2 |{\cal M}(b' \to q l \nu )|^2 
\nonumber \\
      &&  \times~        \frac{1}{(E_X+E_b-E_B)^2} \frac{1}{(2 E_b)^2}
                \frac{1}{(E_g+E_b'-E_b)^2} \frac{1}{(2 E_b')^2}
\nonumber \\
      &&  \times~  (2 \pi)^4 \delta^4(p_B-p_X -q -p_q -p_g)
\nonumber \\
      &&  \times~   \frac{d^3 p_{_X}}{(2\pi)^3 2E_X} 
                   \frac{d^3 p_q}{(2\pi)^3 2 E_q}
                   \frac{d^3 q}{(2\pi)^3 2 q_0}
                   \frac{d^3 p_g}{(2\pi)^3 2 E_g}~,
\ee
and 
\be
d \Gamma(b' \to q W ) = \frac{1}{2 E_{b'}} |{\cal M}(b' \to q W )|^2 
                (2 \pi)^4 \delta^4(p_{b'}-q -p_q )
                   \frac{d^3 p_q}{(2\pi)^3 2 E_q}
                   \frac{d^3 q}{(2\pi)^3 2 q_0}~.
\ee
Substituting Eq. (15) and (22) into Eq. (21), 
we obtain $\Delta f(x)$ from Eq. (18):
\be
\Delta f(x) 
%&=& \frac{m_B}{2 \Gamma_0} 
%\frac{d \Gamma (b' \to q l \nu) + g}{d q_0} 
%\nonumber \\
%&& \times \frac
= \delta (y'-x)~ f(y) dy
 ~\frac{E_b'}{E_b} | {\cal M}(b \to b' g)|^2
                \frac{1}{(E_g+E_b'-E_b)^2} \frac{1}{(2 E_b')^2} 
                   \frac{d^3 p_g}{(2\pi)^3 2 E_g}~.
\ee

Now we need the amplitude for the $b \to b' g$ subprocess 
to calculate $ \Delta f(x) $ and it can be easily calculated.
For simplicity, we take the limit $|{\bf p}_b| >> m_b$
to obtain the Lorentz invariant quantity $ |{\cal M}(b \to b' g)|^2$.
Let the components of the momenta of the $b$ quarks 
before and after emitting a gluon and 
the emitted gluon be:
\be
p_b^\mu &=& (y E_B, {\bf 0}, y E_B)~,
\nonumber \\
p_{b'}^\mu &=& \left(y' E_B + \frac{{\bf k}_T^2}{2y'E_B}, 
             {\bf k}_T, y' E_B \right)~,
\\
p_g^\mu &=& \left( (y-y')E_B  + \frac{{\bf k}_T^2}{2(y-y')E_B}, 
             -{\bf k}_T, (y-y') E_B \right)~,
\nonumber
\ee
where $|{\bf k}_T|$ denotes the size of the transverse momentum 
of emitted gluon.
We also let $z=y'/y$ and obtain the amplitude
\be
|{\cal M}(b \to b' g)|^2 &=& \frac{1}{2} \cdot \frac{1}{3}
        \sum |\bar{u}(p_b) (i g_s \gamma_\mu T^a_{bc}) u(p_b') 
                                         {\epsilon^*}^{\mu}(p_g)|^2
\nonumber \\
 &=& 8 \pi \alpha_s C_2(F) \frac{1+z^2}{z(1-z)^2} |{\bf k}_T|^2~,
\ee
where $ {\epsilon^*}^{\mu}(p_g)$ is the polarization vector 
for gluon and $C_2(F) = \sum_a T^a T^a$ is the casimir of SU(3) group.
The factor $1/2$ in the first line denotes spin average and 
$1/3$ color average.
The change of the momentum transfer arises from the gluon emission
which sets the size of $ |{\bf k}_T|$.
%The momentum transfer $q^2$ is this scale which sets the size of the 
%decaying $b'$ quark transverse momentum $ |{\bf k}_T|$.
%Hence we let $ d (\ln |{\bf k}_T|^2) = d (\ln q^2)$ and
Hence we let $ d |{\bf k}_T|^2/|{\bf k}_T|^2 = d q^2/q^2$ and
we have the contributions to the change of 
the quark distribution function:
\be
\Delta f(x,q^2) 
= \frac{\alpha_s(q^2)}{2 \pi} \int_0^1 dy~ dz~ f(y,q^2)~ P(z)~ 
 [ \delta(yz - x)-\delta(y-x)]~ d (\ln q^2)~,
\ee
where 
\be
P(z) = C_2(F) \frac{1+z^2}{1-z}
\ee
represents the probability for a quark radiating
a gluon such that the quark's momentum is reduced by a fraction $z$
and has the same form as the non-singlet contribution
of the splitting function of DIS.
Note that we should have the negative contributions to $\Delta f(x)$
as well as the positive one.
The negative contribution arises when the $b$ quark originally 
has the momentum fraction $x$ and the gluon emission reduces 
the momentum fraction into $y'=zy$.
While the positive contribution term $\delta(y'-x)$ implies 
production of a quark with momentum fraction $x$, 
the negative contribution term $\delta(y-x)$ implies 
elimination of a quark with momentum fraction $x$.

Finally we write the evolution equation which is the same form 
as the non-singlet contribution of the Altarelli--Parisi equation:
\be
\frac{df(x,t)}{dt} = \frac{\alpha_s(t)}{2 \pi} \int \frac{dy}{y} 
             ~f(y,t)~ P \left( \frac{x}{y} \right)~,
\ee
where $t = \frac{1}{2} \ln q^2/q^2_0$ and
\be
P(z) = C_2(F) 
  \left[ \frac{(1+z^2)}{(1-z)_+} + \frac{3}{2} \delta(1-z) \right]~,
\ee
with the notation of the function regularization 
\be
\int_0^1 dz \frac{f(z)}{(1-z)_+} = \int_0^1 dz \frac{f(z)-f(1)}{1-z} ~,
\ee
which denotes the regularization of the singularity due to soft gluon.
This is introduced because this integral actually diverges as $z \to 1$.
We should also have included the vertex correction diagram for the
infrared divergence cancellation
when we squared the one--gluon emission diagram.

\section{Evolution of moments}

We define the moments of the distribution function
$f(z)$ by
\be
M_n(t) = \int_0^1 dz~z^n~f(z,t)~,
\ee
which yields the evolution equation of moments
\be
\frac{d M_n(t)}{dt} = \frac{\alpha_s(t)}{2\pi} M_n(t) D_n~,
\ee
where 
\be
D_n = \int_0^1 dz ~z^n P(z)~.
\ee
The evolution is caused by running of the strong coupling constant 
$\alpha_s(t)$ and Eq. (32) yields
\be
\frac{d M_n}{M_n} = -\frac{D_n}{(11-\frac{2}{3}n_f)} 
                           \frac{d \alpha_s}{\alpha_s}~.
\ee
We obtain the solution for evolution of moments:
\be
\frac{M_n(t)}{M_n(0)} =  \left( 
             \frac{\alpha_s(t)}
                  {\alpha_s(0)} \right)^{-\frac{D_n}
                                             {(11-\frac{2}{3}n_f)}} ~.
\ee
If we have an model distribution function at $q^2 = q^2_0$
as an initial condition, its moments $M_n(0)$ can be explicitly
calculated and we obtain the evolved moments $M_n(t)$ directly.

To show the result in terms of the HQET parameters,
we consider the relations of moments of the light--cone
distribution function $F(x)$ defined by the HQET and of our
distribution function defined in Eq. (7).
These functions are related by a change of variable
\be
\frac{1}{m_B} f(x_+) = \frac{1}{\bar{\Lambda}} F(x_B)~,
\ee
where
\be
x_B = - \frac{m_b^2+q^2-2 m_b q_0}
             {2 \bar{\Lambda} (m_b-q_0)}~.
\ee

If we define the moments of $F(x_B)$ as
\be
a_n = \int dx_B~x_B^n~F(x_B)~,
\ee
the first moment is independent of the scale 
according to the normalization of the function
\be
a_0 = 1 = M_0(t)~,
\ee
which tells us the $b$--number conservation
corresponding to the moments sum rule of the DIS process.

The second moment $a_1$ is equal to be zero by the equation
of motion of the HQET without running, 
which indicates the absence of 
the leading correction in $1/m_b$ expansion.
It holds when we take the pole mass of the $b$ quark 
as the expansion parameter
and is related to the definition of the $b$ quark mass.
In Ref. \cite{kylee1} 
it is shown that the parton model formulation
can have this property if we define the $b$ quark mass by the Eq. (1).
Including running effects, however, Eq. (1) no more holds
and the $b$ quark mass should be redefined:
\be
m_b = m_{_B} M_1(t) \left( \frac{\alpha_s(t)}{\alpha_s(0)}
                  \right)^{\frac{D_1}{11-\frac{2}{3} n_f}}~,
\ee
which results in
\be
a_1(t) = \frac{m_{_B}}{\bar{\Lambda}} M_1(t)
\left( 1 - \left( \frac{\alpha_s(t)}{\alpha_s(0)}
                  \right)^{\frac{D_1}{11-\frac{2}{3} n_f}} \right).
\ee
We can compute it using $M_1(t)$ and $D_1$.

The third moment $a_2$ is related to the average kinetic energy
of the $b$ quark inside $B$ meson, $\mu_\pi^2$. 
So the running  effects is expressed by the running of $\mu_\pi^2$ 
and computed by
\be
\mu_\pi^2(t) = 3 m_{_B}^2 (M_2(t) - M_1^2(t))~.
\ee

\section{Solving the equation and phenomenologies}

Since our evolution equation has the same form as the non-singlet
part of the Altarelli--Parisi equation for the DIS, we use the
%The Altarelli-Parisi equation is numerically solved using the
package developed by Miyama, et al.\cite{bf1}
for solving the Altarelli--Parisi equation numerically.
This program adopted, so-called, the brute--force method
which is the simplest one in solving the integrodifferential equation.
It divides the variable $x$ and $t$ into small steps
and defines differentiation as $df(x)/dx = (f(x_{m+1})-f(x_m))/\Delta x_m$
and integration as $\int dx f(x) = \sum^N_{m=1} \Delta x_m f(x_m)$.
Thus the evolution equation could be solved by repeated 
summations and subtractions.
This method provides us the numerical accuracy as high as we want 
by spending running time enough if the initial distribution function
$f(x,q^2_0)$ is given.
At the low energy scale $q^2 \sim \Lambda_{QCD}$, the nonperturbative 
effects dominate and it is reasonable to take a modelling
of the distribution function as an initial condition
for the evolution at this scale.
Here we use the fragmentation inspired form suggested by
Lee and Kim \cite{kylee1} as the initial condition at 
$q_0^2=1$ GeV$^2$ 
which improves Peterson's fragmentation function
\be
  f(x,q^2=q_0^2) = \frac{N}
          {x \left(1+\alpha-{1\over x}-{\epsilon \over 1-x} \right)^2}~.
\ee
For numerical analysis, we took the values $\alpha_Q=0.085$, 
$\epsilon_Q=0.004$.
In this case, the normalization constant is fixed as
$N_Q =0.0235$ by the condition
\be
  \int_0^1 dx~ f(x,q^2) =1.
\ee
We solve the equation in the leading order of $\alpha_s$.

The distribution functions for varying $q^2$ are
shown in Fig.2.
We can see that it becomes rapidly broader as $q^2$ increases, which 
suggest the QCD leading order resummation is large. 
We also show the plots of $q^2$-dependences of the distribution function
for different values of $x$ in Fig. 3.
We can understand these figures as follows:
If $x$ is large, the $b$ quark has large momentum and would
radiate more energetic gluons,
which reduce the momentum of $b$ quark.
As a result, the low $x$ region receives 
much positive contribution from the large $x$ region 
and only a little negative contribution,
while in the region around the peak, 
negative contribution dominates and 
positive contribution are very little.

On the other hand, the observable such as $E_e$ distribution
is not so much affected as shown in Fig. 4
because of the kinematical factor.
We find that $q^2$ spectrum reflects the evolution
effects of the distribution function more directly. 
It is recommended to measure the $q^2$ spectrum accurately
for probing the evolution effects from perturbative QCD.

\section{Conclusion}

We derived and solved the evolution equation
of the light--cone distribution function of a $b$ quark
inside $B$ meson to combine the perturbative QCD
corrections for inclusive semileptonic decays.
In the viewpoint of parton model, the nonperturbative
effects are encoded in the structure function of the decay process
which discribes the distribution of $b$ quark momentum.
Since $b$ quark is heavy enough, the momentum transfer
of decay is larger than the QCD scale in most region of the phase space.
It enables us to treat the $b$ quark as a noninteracting parton.

With the scaling property set in,
the leading correction of perturbative QCD comes from 
the one gluon emission.
The evolution of the distribution function is obtained by
adding the one gluon emission diagram which violates
the scaling behavior of inclusive decays.

The distribution function obeys kinds of sum rules
since the moments of the function can be directly related
to the HQET parameters.
If the mass of $b$ quark is properly defined in the model framework,
we can apply the $1/m_b$ expansion technique to the model study
and that mass should be the pole mass.
When the evolution effects comes into consideration,
the pole mass is redefined to modify the equation of motion
by the power corrections \cite{balzereit}.
The fact that the second moment of the genuine QCD
distribution function $F(x)$ is equal to zero
defines the $b$ quark mass $m_b$ by Eq. (1)
from the relation
\be
a_1 = \frac{m_{_B}}{\bar{\Lambda}} 
      \left( M_1 - \frac{m_b}{m_{_B}} \right) = 0~.
\ee
If $a_1$ is kept to be zero with running,
the value of $m_b(\mu) = M_1(\mu) m_{_B}$ changes
and this $m_b$ cannot be the pole mass.
Therefore we should modify the definition of the pole mass
to compensate the running effects as in Eq. (40)
which indicates that $a_1$ is no more zero with evolution effects.
In our approach, we explicitly present that the definition of
mass is changed by the running of $a_1(\mu)$.

We numerically solve the equation
and the charged lepton energy and $q^2$ spectra are also shown.
The shapes of decay spectra are changed by incorporating
the perturbative corrections, while charged lepton energy
spectrum is less sensitive.
The decay rate decreases with the corrections and
we conclude that
the perturbative corrections should be considered
in practically extracting the CKM parameters.

\vskip 0.8cm
%\acknowledgement
\begin{center}
{\bf Acknowledgement}
\end{center}
\vskip 0.6cm

KYL thanks to Professor Junegone Chay for helpful discussions.
This work is supported by the Korean Science and Engineering Foundation 
(KOSEF) through the SRC program of the Center for Theoretical Physics (CTP)
at Seoul National University.

%\newpage
%{\Large \bf Figure Captions}
%\vskip 2cm
\begin{figure}
\caption{
(a) Diagram of inclusive semileptonic decay of $B$ meson.
\\
(b) One gluon emission diagram in Inclusive semileptonic 
decay of $B$ meson.
}
\end{figure}

\begin{figure}
\caption{
The light-cone distribution functions of $b$ quark 
for $q^2 =1.0$ (solid), 5.0 (dashed), 25.0 (dash-dotted) GeV$^2$.
The curve for $q^2 =1.0$ GeV$^2$ is the initial distribution.
}
\end{figure}

\begin{figure}
\caption{
The evolutions of the light-cone distribution functions for fixed $x$.
From up to down $x=0.9$, 0.8, 0.7, 0.5, 0.1.
}
\end{figure}

\begin{figure}
\caption{
The $q^2$-spectra of inclusive semileptonic $B$ decays:
The solid (dashed) line denotes the case
without (with) the QCD evolution effect.
%that the QCD evolution effect is (not) considered.
Only the $q^2 > 1.0$ GeV$^2$ region from which the distribution
function is evolved is shown in the figure.
}
\end{figure}

\begin{figure}
\caption{
The electron energy spectra of inclusive semileptonic $B$ decays:
The solid (dashed) line denotes the case
without (with) the QCD evolution effect.
}
\end{figure}

\newpage
\setcounter{figure}{0}

\begin{figure}[th]
%\caption{}
\centering
\centerline{\epsfig{file=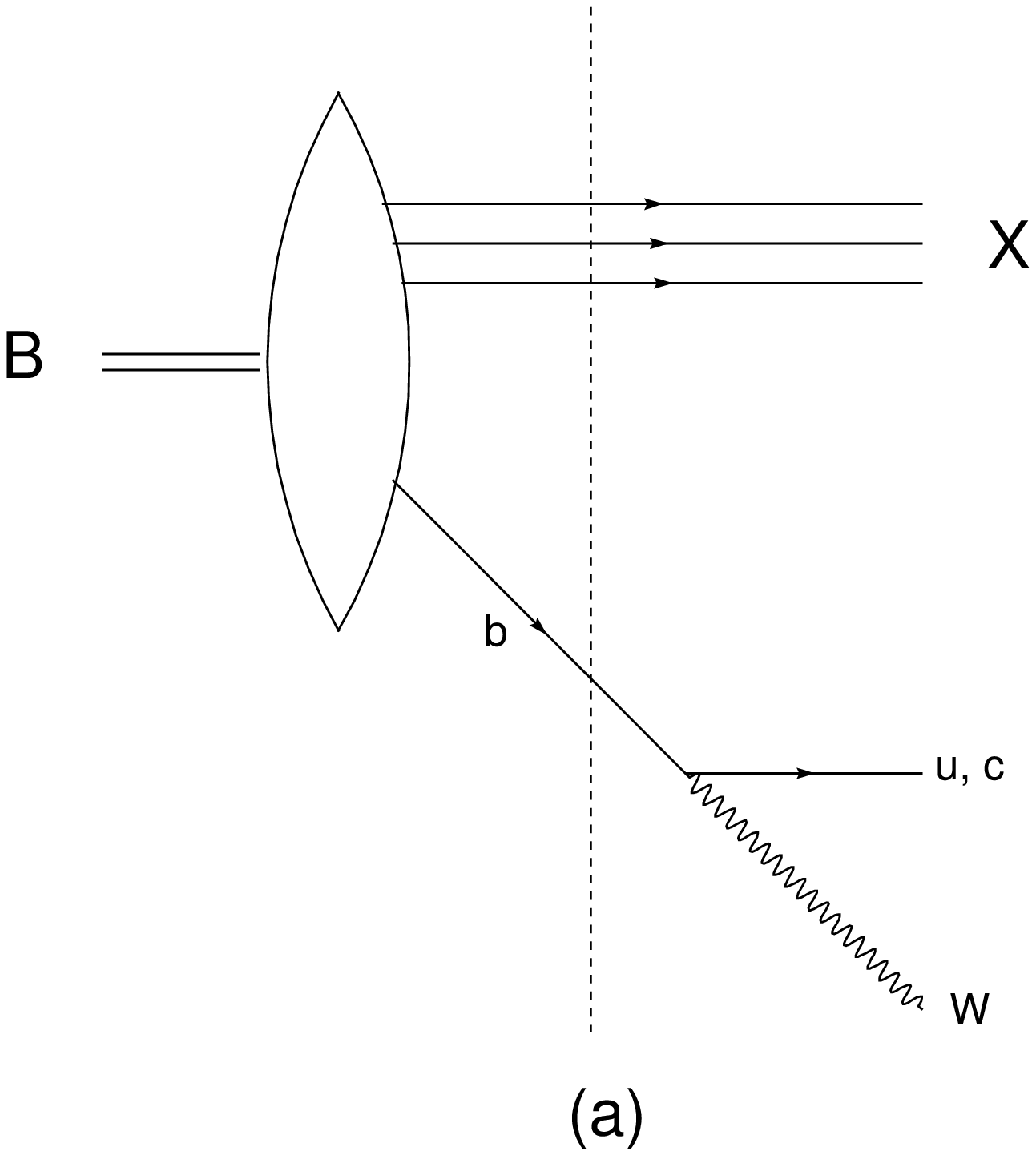}}
\end{figure}
\begin{center}
{\large Fig. 1}
\end{center}

\setcounter{figure}{0}

\begin{figure}[th]
%caption{}
\centering
\centerline{\epsfig{file=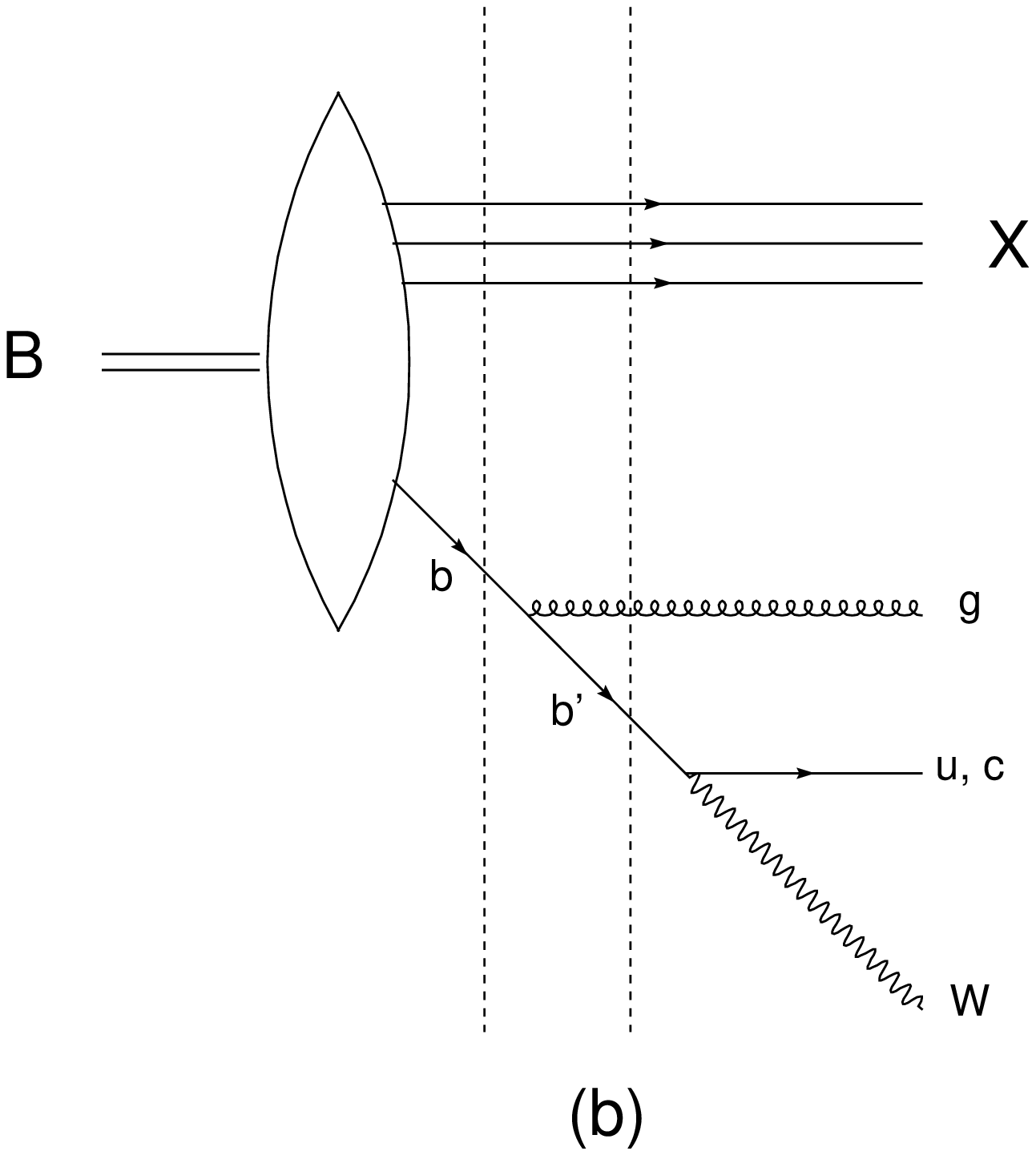}}
\end{figure}
\begin{center}
{\large Fig. 1}
\end{center}

\begin{figure}[th]
%caption{}
\centering
\centerline{\epsfig{file=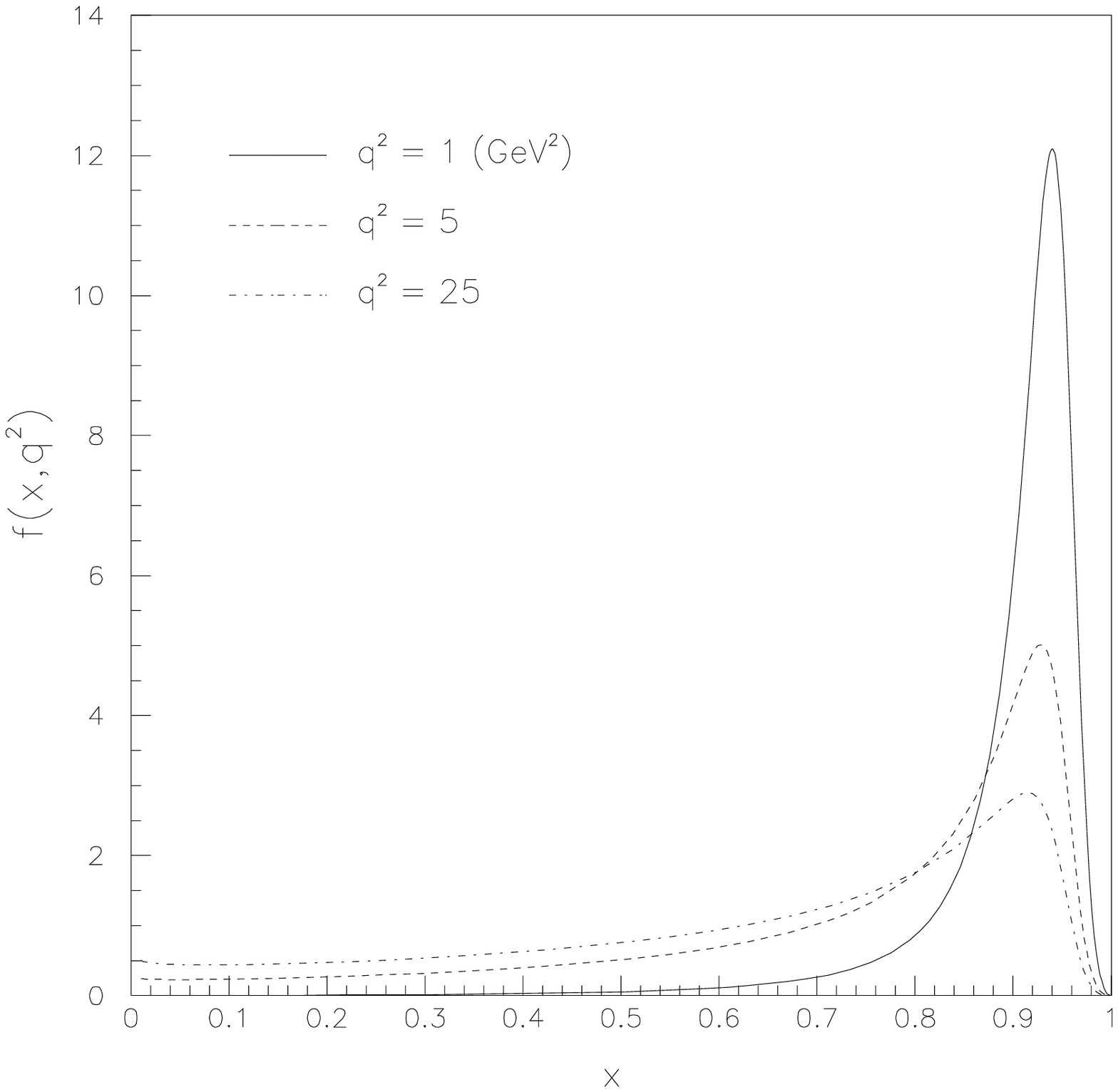}}
\end{figure}
\begin{center}
{\large Fig. 2}
\end{center}

\begin{figure}[th]
%caption{}
\centering
\centerline{\epsfig{file=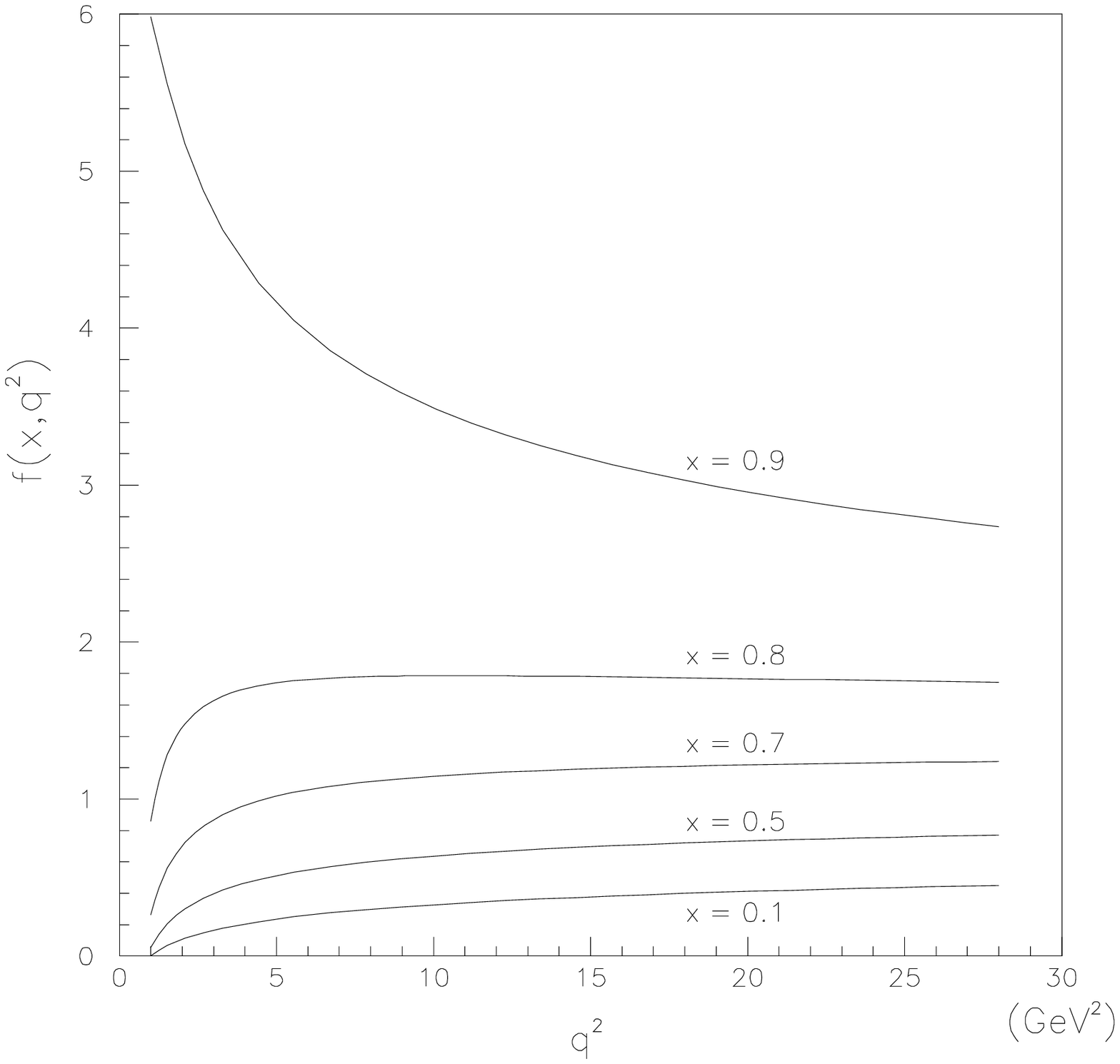}}
\end{figure}
\begin{center}
{\large Fig. 3}
\end{center}

\begin{figure}[th]
%caption{}
\centering
\centerline{\epsfig{file=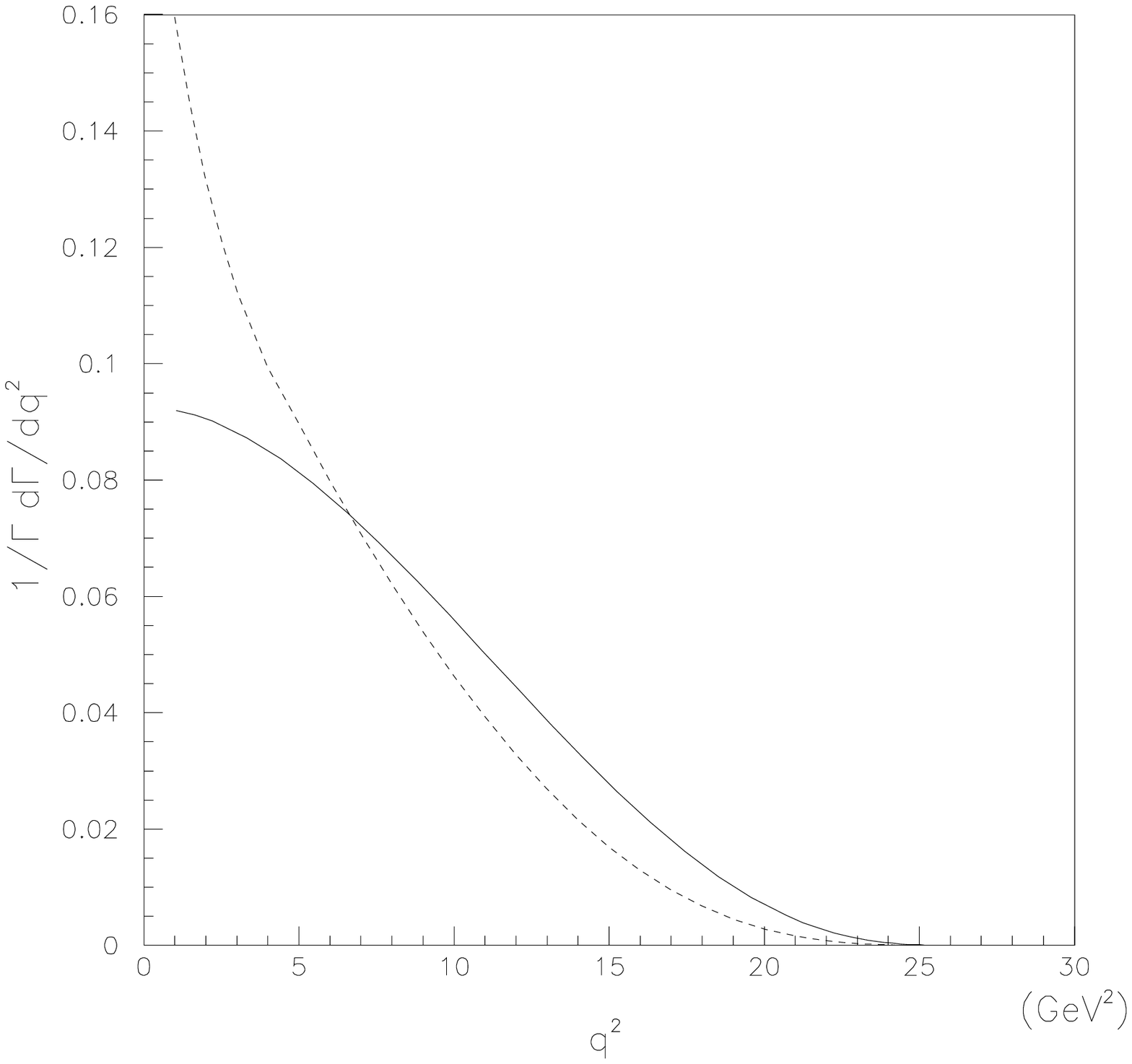}}
\end{figure}
\begin{center}
{\large Fig. 4}
\end{center}

\begin{figure}[th]
%caption{}
\centering
\centerline{\epsfig{file=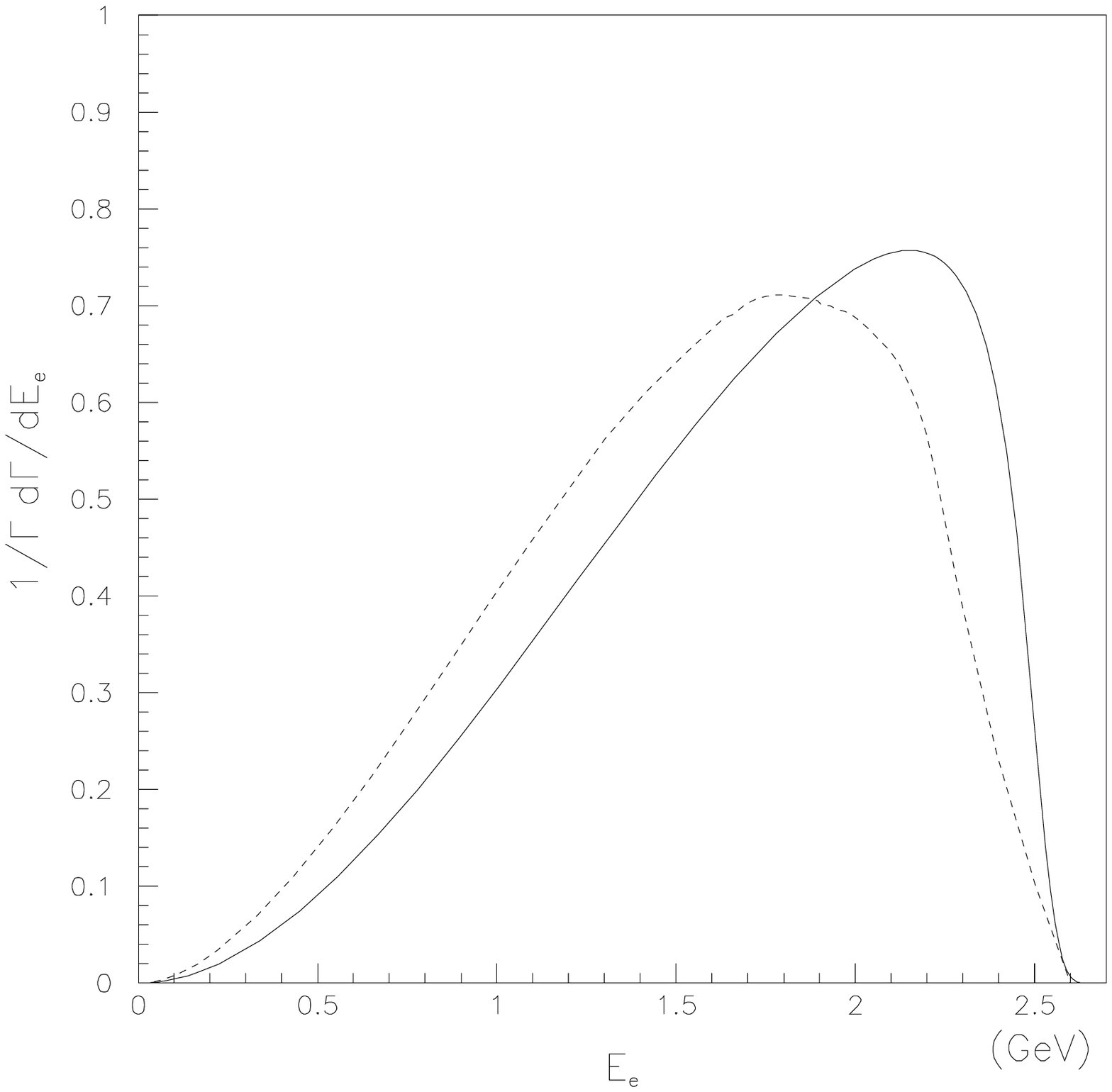}}
\end{figure}
\begin{center}
{\large Fig. 5}
\end{center}

\end{document}